\documentclass{emulateapj}

\usepackage{amsmath}
\usepackage{color}

\slugcomment{}

\shorttitle{Double-Sided Jet in GRB 170817A}
\shortauthors{Li, Geng, Huang \& Li}

\begin{document}


\title{Late-Time Afterglow from Double-Sided Structured Jets: Application to GRB 170817A}


\author{Long-Biao Li\altaffilmark{1,3}, Jin-Jun Geng\altaffilmark{2,3}, Yong-Feng Huang\altaffilmark{2,3}, Bing Li\altaffilmark{2,4,5}}


\altaffiltext{1}{School of Mathematics and Physics, Hebei University of Engineering, Handan 056005, People's Republic of China; lilongbiao@hebeu.edu.cn}
\altaffiltext{2}{School of Astronomy and Space Science, Nanjing University, Nanjing 210023, People's Republic of China; gengjinjun@nju.edu.cn}
\altaffiltext{3}{Key Laboratory of Modern Astronomy and Astrophysics (Nanjing University), Ministry of Education, Nanjing 210023, People's Republic of China}
\altaffiltext{4}{Key Laboratory of Particle Astrophysics, Chinese Academy of Sciences, Beijing 100049, People's Republic of China}
\altaffiltext{5}{Laboratory for Particle Astrophysics, Institute of High Energy Physics, Beijing 100049, People's Republic of China}


\begin{abstract}
The broadband afterglow of GW170817/GRB 170817A is believed to be from an off-axis structured jet.
The central engine of a gamma-ray burst usually launches a pair of outflows that move oppositely,
it is reasonable to consider the emission from a double-sided structured jets,
with a near-jet moving toward us and a counter-jet moving away from us.
Assuming that the two branches of the jet have the same physical parameters,
we have calculated their radio emission.
It is found that the counter-jet component will emerge in the radio light curves
$\sim$ 2500 days post-merger.
It typically leads to a plateau in the light curve,
thus could be hopefully revealed by high accuracy radio observations.
We have also considered the possibilities that both branches have
different parameters, and found that if some of the physical parameters
of the counter-jet are evaluated more favorably than
those of the near-jet, then the counter-jet emission will be
enhanced and will clearly show up as an obvious plateau or even a rebrightening.
For example, when the circum-burst medium encountered by the counter-jet is
assumed to be 100 times denser than that of the near-jet,
a remarkable radio plateau will appear at $\sim$ 600 days.
However, in X-ray bands, the counter-jet component is generally too faint to be discerned.
It is argued that the late radio observations of GW170817/GRB 170817A
can help determine the key parameters and diagnose the environment of the event.
\end{abstract}


\keywords{gamma-ray burst: individual: GRB 170817A --- ISM: jets and outflows --- methods: numerical}

\section{Introduction}

The first gravitational wave (GW) signal from a binary neutron star (NS) merger, GW170817, which was detected
by the advanced Laser Interferometer Gravitational Wave Observatory (LIGO) and the Virgo Interferometer \citep{Abbott2017a},
was followed by a faint short gamma-ray burst (GRB) GRB 170817A
\citep[e.g.,][]{Abbott2017b,Goldstein2017,Savchenko2017},
a $r$-process-induced kilonova \citep[e.g.,][]{Arcavi2017,coulter2017,Covino2017,drout2017,Pian2017,Smartt2017},
and a long-time broadband (e.g., radio, optical, X-ray) afterglow \citep[e.g.,][]{evans2017,Hallinan2017,
Margutti2017,troja2017,DAvanzo2018,Lamb2018,lyman2018,mooley2018a,Mooley2018,piro2018MN,Ruan2018,Troja2018}.

The low luminosity of GRB 170817A and the non-detection of early-time ($\lesssim$ a few days) afterglow imply that
GRB 170817A should not be a typical short GRB with an on-axis line of sight
\citep[e.g.,][]{Kasliwal2017,Kathirgamaraju2018,Meng2018}.
The subsequent temporal evolution of the multi-wavelength emission,
which is characterized by an initial steady shallow rise ($\propto t^{0.9}$) in tens of days
and a steep decline ($\propto t^{-2}$) beyond $\sim$ 150 days post-merger
\citep{Alexander2018,Dobie2018,Lamb2018,mooley2018a,piro2018MN,Troja2018,vanEerten2018},
confirmed the presence of an energetic off-axis jet,
but ruled out the uniform ``top-hat'' jet structure
\citep[e.g.,][]{Alexander2017,Haggard2017,Margutti2017,
Murguia2017,troja2017,Xiao2017}.
Similarly, a mildly relativistic isotropic cocoon with a choked jet is also not preferred
\citep[e.g.,][]{Abbott2017c,Lazzati2017b,Murguia2017,Piro2018}.
Instead, the continuously-rising emission could be interpreted by
a number of scenarios, including
a continued injection of energy from the central engine into the external jet
\citep{geng2018,lib2018,pooley2018},
interaction of the dynamic ejecta tails of the merger ejecta with the surrounding medium
\citep{Hoto2018},
a structured jet with a highly relativistic inner core, or a jet-cocoon system
produced in the NS merger
\citep[][etc.]{DAvanzo2018,Lazzati2018,LK2018,lyman2018,Margutti2018,Mooley2018,
Nakar2018,Troja2018,Xie2018,Geng2019}.
However, the turn-over in the light curve at $\sim$ 150 days post-merger
and the ongoing rapid decline disfavor most cocoon systems, and are
consistent with the emergence of a relativistic off-axis structured jet
\citep[e.g.,][]{Lamb2018a,Lazzati2018,mooley2018a,piro2018MN}.

It is worth mentioning that the central engines of GRBs (either accreting neutron stars or
accreting black holes) should launch a pair of jets in principle.
One branch of the jets, moving toward the observer, is called the near-jet.
It produces the normal prompt emission and multi-wavelength afterglow of the GRB.
The other branch, called the counter-jet, is moving away from the observer.
Initially, since the counter-jet is also ultra-relativistic and its emission is
highly beamed, it is almost completely invisible for us.
However, when the counter-jet slows down and becomes non-relativistic,
its emission will be sent to a wider and wider angular range and finally will be
nearly isotropic. It will then be also visible for us and may lead to a short plateau or even a
rebrightening in the late-time afterglow light curve
\citep{Granot2003,Li2004,Zhang2009,Wang2009,Wang2010,WangH2010,vanEerten2011}.
The counter-jet has been mentioned in regards to GRB 170817A by \citet{Gill2018} and \citet{Lamb2018}.
They stated that the counter-jet starts to contribute in the afterglow light curve at $>$ 1000 days,
and peaks/dominates at $\sim 10^4$ days.
Besides, \citet{Zrake2018} have simulated radio sky-maps of GW170817A/GRB 170817A and shown
the appearance of the counter-jet at $\sim 800$---1000 days.
In some extreme cases, it is argued that the emission from the counter-jet
may even be detectable at about 10 --- $10^3$ s post-merger \citep{Yamazaki2018}.
Nevertheless, the radiation from the counter-jet is generally ignored
by most researchers, since this component is usually very weak and thus has not
been detected yet \citep{Wang2010,Yamazaki2018}.

However, GW170817/GRB 170817A provides us a valuable opportunity for studying
the counter-jet emission. There are two reasons for this. First, it is an event
with an off-axis jet geometry. In this case, the emission from the near-jet
is significantly reduced while the emission from the counter-jet is strongly
enhanced, which ensures us more likely to succeed in detecting the counter-jet.
Second, GW170817/GRB 170817A is not too far away from us, so that its late time
afterglow could be continuously monitored. This is very important in revealing the
counter-jet component \citep{Zhang2009,Yamazaki2018}.
In this article, we will present our detailed numerical study on the emission from
double-sided jets in both relativistic and Newtonian stages. We assume that each
branch of the jets is a structured outflow to mimic GRB 170817A.
We compare our results with the observations of this famous binary neutron star merger event.
The structure of our article is as follows.
In Section 2, we describe the dynamical evolution of the structured near-jet and
counter-jet briefly.
In Section 3, numerical results on the afterglow of the double-sided jets are presented and
compared with observations.
Finally, Section 4 is our discussion and conclusion.

\section{Double-Sided Structured Jet}
\label{jet}

A structured jet, different from a uniform ``top-hat'' jet,
is characterized by a narrow, highly relativistic inner core,
surrounded by some less energetic, slower wings at larger angles
\citep[e.g.,][]{Dai2001,Lipunov2001,Rossi2002,Zhang2002,Kumar2003}.
Before the relativistic outflow launched by the central engine
gives birth to a successful GRB, it needs to push through
a significant amount of material ejected by the progenitor star
\citep{Bromberg2011,Nagakura2014,Nakar2017}.
The structure of the jet thus can either be due to the jet formation mechanism
itself \citep{van2003,Vlahakis2003,Aloy2005}, or can be
resulted from the breaking out process in the stellar envelope
\citep{Levinson2003,Zhang2003,Lazzati2005,Morsony2010,Pescalli2015}.

After the structured jet produced the prompt emission like GRB 170817A
\citep{Kasliwal2017,Lazzati2017a,Lazzati2017b,Meng2018}, it continues to move outward
and excites an external shock propagating into the interstellar medium \citep{Jin2018,LK2018}.
The interaction between the shock-accelerated electrons and magnetic field
produces synchrotron radiation, giving rise to the broadband afterglow emission
\citep{M1997,Sari1998,SPN1998,Zhang2004,Zhang2014}.

Within a structured jet, the kinetic energy per solid angle $(\varepsilon)$
and bulk Lorentz factor $(\Gamma)$ should vary with angle from the central axis.
Here, we assume that the structured jet has a uniform central core and
its overall profile could be described as \citep{Dai2001,Lipunov2001,Rossi2002,Zhang2002,Kumar2003}
\begin{equation}
\varepsilon(\theta) = \left\{
             \begin{array}{ll}
             {\varepsilon_c,}  &\theta < \theta_c, \\
             {\varepsilon_c(\theta/\theta_c)^{-k_e},}  &\theta_c < \theta < \theta_m,
             \end{array}
        \right.
\end{equation}
\begin{equation}
\Gamma(\theta) = \left\{
             \begin{array}{ll}
             {\Gamma_c,}  &\theta < \theta_c, \\
             {\Gamma_c(\theta/\theta_c)^{-k_\Gamma}+1,}  &\theta_c < \theta < \theta_m,
             \end{array}
        \right.
\end{equation}
where $\theta_c$, $\varepsilon_c$ and $\Gamma_c$ are respectively the half-opening angle,
the kinetic energy density and Lorentz factor of the inner core,
$\theta_m$ is the maximum half-opening angle of the jet, and the indexes $k_e$ and $k_\Gamma$
describe the angular distribution of $\varepsilon(\theta)$ and $\Gamma(\theta)$,
respectively.

For the double-sided jet considered here, we split the structured outflow of both the
near-jet branch and the counter-jet branch into $10^4$ small patches (i.e., 100 segments
along both $\theta$ and $\phi$ directions).
The dynamics and synchrotron radiation are calculated for each patch separately.
Then, the total afterglow emission is calculated by summing up the emission from all the patches.
For the counter-jet branch, the relativistic beaming effect should be considered, which strikingly
reduces the emission frequency and intensity in the relativistic phase. Additionally, the
observer's time is delayed by a period of $2r/c$
due to the light-travel effect, where $r$ is the radius of the corresponding patch,
and $c$ is the speed of light \citep{Li2004}.
As a result, the counter-jet branch emission should be calculated by
$F_{\nu, {\rm cj}}(t)=F_{\nu, {\rm nj}}(t-2r/c)$,
where, here and hereafter, the subscript cj and nj indicate respectively the counter-jet and near-jet.
In our calculations, each patch is assumed to be independent of others for simplicity.
The dynamical evolution of each patch is calculated numerically
by adopting the generic dynamical equations suggested by Huang et al.
\citep{Huang1998,Huang1999a,Huang1999b,Huang2000a,Huang2000b},
which are valid in both ultra-relativistic and non-relativistic stages,
and can be widely applied to calculate the overall afterglow light curves
under various physical conditions
\citep{Wu2004,Huang2006,Kong2010,Yu2013,Geng2013,Geng2014,Li2015MN,Li2015A,Zhang2015,lib2018}.
In addition, the effect of equal arrival time surfaces \citep{Waxman1997,Sari1998,Huang2000a,Huang2000b} is also
taken into account.

\begin{figure*}[ht!]
\centering
\includegraphics[width=1.0\textwidth]{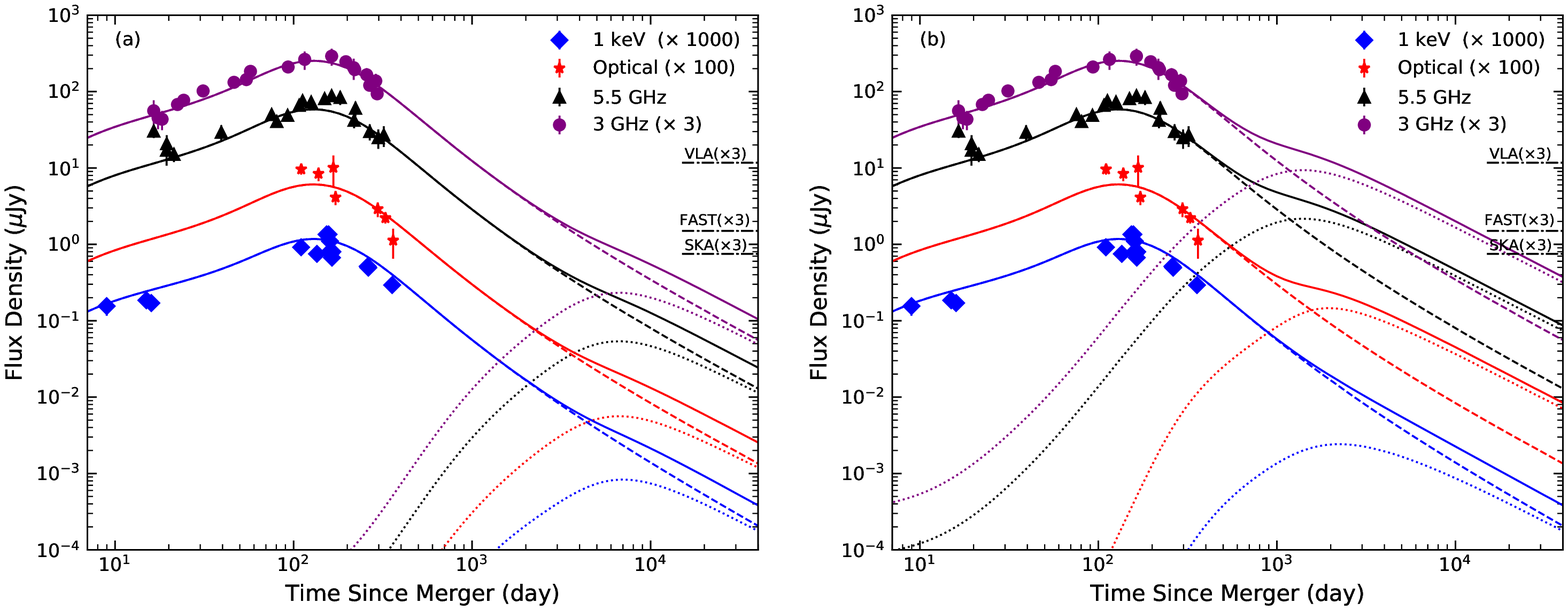}
\caption{
Our theoretical afterglow light curves as compared with the multi-wavelength
datasets of GRB 170817A.
In each panel, the dashed lines correspond to the contribution from the near-jet
and the dotted lines correspond to the contribution from the counter-jet.
The solid lines show the total emission from the double-sided structured jet.
In the left panel, the same set of parameters are applied to both the counter-jet
and the near-jet. In the right panel, the number density of the ambient medium
of the counter-jet is assumed to be 100 times higher,
i.e., $n_{\rm cj}$=100 $n_{\rm nj}$, while other parameters are unchanged.
The observed radio data points at 3 GHz and 5.5 GHz are compiled from
\citet{Alexander2017}, \citet{Alexander2018}, \citet{Hallinan2017}, \citet{Kim2017},
\citet{Dobie2018}, \citet{Margutti2018}, \citet{Mooley2018},
\citet{mooley2018a}, \citet{piro2018MN}, and \citet{vanEerten2018}.
The observed optical data points at $5.1 \times  10^{14}$ Hz (F606W) are adopted from
\citet{Lamb2018}, \citet{lyman2018}, \citet{Margutti2018} and \citet{piro2018MN}.
The observed X-ray data points are taken from \citet{Alexander2018}, \citet{DAvanzo2018},
\citet{Lazzati2018}, \citet{piro2018MN}, \citet{Ruan2018} and \citet{vanEerten2018}.
The horizontal solid bars represent the 1$\sigma$
sensitivities (assuming an integration time of one hour) of FAST, SKA and VLA.
}
\label{fig1}
\end{figure*}

\begin{figure}[ht!]
\includegraphics[width=0.5\textwidth]{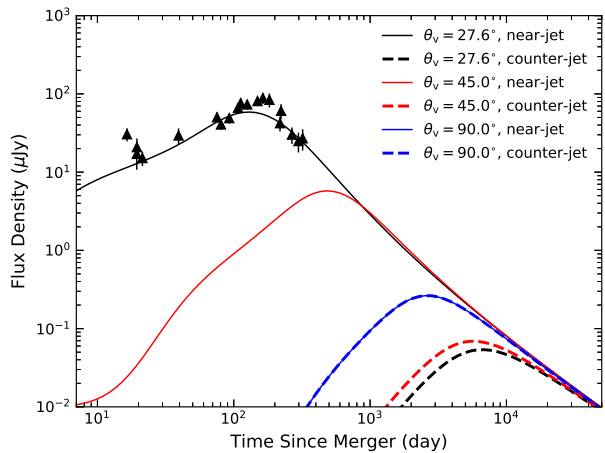}
\caption{Theoretical afterglow light curves of a double-sided structured jet at 5.5 GHz, showing
the effects of the observing angle.
In this illustration, the twin jets are assumed to have the same physical parameters.
The solid, dashed and dotted lines correspond to the cases of $\theta_{\rm v} = 27.6 ^{\circ}$,
$45.0 ^{\circ}$, and $90.0 ^{\circ}$, respectively.
For each viewing angle, the near-jet emission is plotted by the thin solid line,
and the counter-jet component is presented by the thick dashed line, respectively.
Note that for the case of $\theta_{\rm v} = 90.0^{\circ}$, the emission from the two branches are actually the same,
so that the blue solid line and the blue dashed line are superposed.
The source of the observed data points are the same as those in Figure \ref{fig1}.
}
\label{fig2}
\end{figure}

\section{Application to GRB 170817A}

We have considered the dynamical evolution of double-sided structured jets,
and calculated the overall afterglow light curves. In this section, we
compare our results with the observed multi-wavelength afterglow of GW170817/GRB 170817A,
which has been monitored for more than one year \citep{Lamb2018,mooley2018a,piro2018MN,Troja2018}.

In our study, we first consider the case that the counter-jet has the same physical parameters
as that of the near-jet. The derived afterglow light curves
are shown in the left panel of Figure \ref{fig1}.
In our calculations, the half-opening angle of the inner core is adopted as
$\theta_c = 5.1^{\circ}$, with an isotropic equivalent kinetic energy
of $E_{c} = 3.4 \times 10^{51}$ erg.
The viewing angle between the near-jet axis and the line of sight
is taken as $\theta_{\rm v}=27.6^\circ$.
The key micro-physical parameters characterizing the energy fraction of electrons and magnetic field
are evaluated as $\epsilon_{e}=0.02$ and $\epsilon_{B}=0.003$, respectively.
The number density of the ambient medium is set as $n=3.3 \times 10^{-3}\,\rm cm^{-3}$.
The above parameters are consistent with those derived in a few recent studies on this
binary neutron star merger event \citep[e.g.,][]{Hallinan2017,Lazzati2018}.
Meanwhile, we adopt the initial Lorentz factor of the inner core as $\Gamma_c =100$.
For the electron spectrum index $p$, it is fixed as $p=2.17$
\citep[e.g.,][]{Margutti2018,vanEerten2018}.
In addition, the index values for the power-law structure of the jets are set as $k_e =4.3$
and $k_\Gamma=2.0$.

\begin{figure*}[ht!]
\centering
\includegraphics[width=1.0\textwidth]{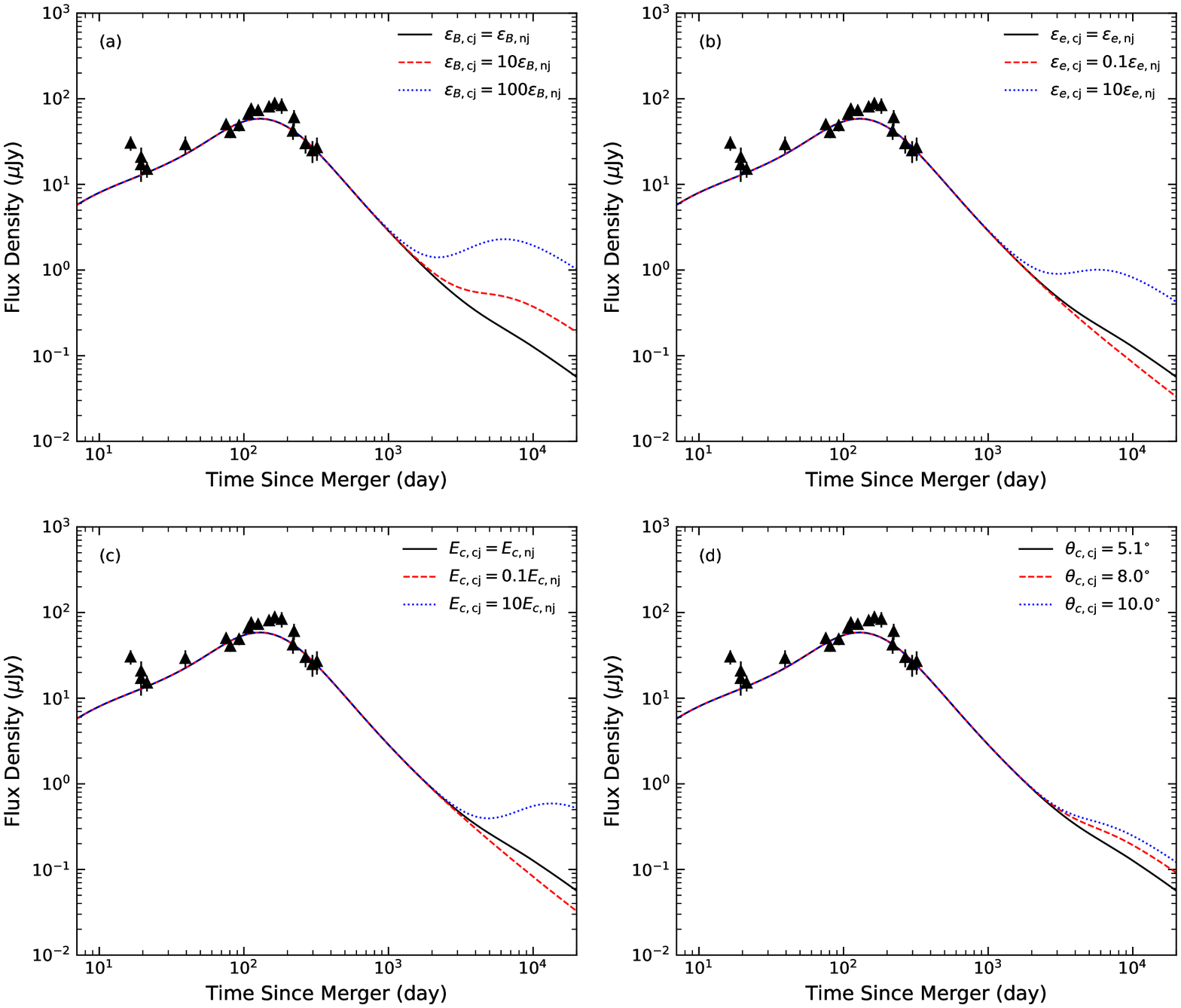}
\caption{Theoretical afterglow light curves of a double-sided structured jet at 5.5 GHz,
showing the effects when various parameters are different for the near-jet and the
counter-jet. Contribution from the counter-jet
is included in the calculated afterglow light curves.
In each panel, the solid line illustrates the case that
the physical parameters are the same for the twin jets. For the detailed parameter
values, please see those in Figure~(1a).
The dashed and dotted lines are plotted with only one parameter being changed for
the counter-jet. In the four panels, the altered parameters are $\epsilon_{B,\rm cj}$,
$\epsilon_{e,\rm cj}$, $E_{c,\rm cj}$, and $\theta_{c,\rm cj}$, respectively.
The source of the observed data points are the same as those in Figure \ref{fig1}.
}
\label{fig3}
\end{figure*}

The left panel of Figure \ref{fig1} illustrates the case
when the physical parameters are the same for the two branches of the double-sided jet.
\citet{Zhang2009} have estimated analytically
the time when the contribution from the counter-jet emerges in the total light curve as
$t_{\rm em,cj}=1900 (1+z) (E_{\rm cj,51} n_{\rm cj,-2})^{-1/3}$ days,
where $E_{\rm cj,51}$ is the isotropic energy of the counter-jet in units of $10^{51}\,\rm erg$,
and $n_{\rm cj,-2}$ is the circum-burst density in units of $10^{-2}\,\rm cm^{-3}$.
Substituting our parameters of the central inner core,
which carries most of the jet's kinetic energy, into this expression,
we can estimate $t_{\rm em,cj} \approx 4.2 \times 10^{3}$ days.
This value, as stated by \citet{Zhang2009}, would be overestimated for the case of off-axis observation.
In fact, the two-dimensional hydrodynamic numerical simulations by \citet{Zhang2009}
indicate that the counter-jet component emerges at $\sim t_{\rm em,cj}/2 \approx 2.1 \times 10^{3}$ days.
Our numerical results in Figure (\ref{fig1}a) show that the counter-jet emission
begins to contribute significantly in the total radio afterglow light curve at about 2500 days post-merger,
which is roughly consistent with that of \citet{Zhang2009}.
In the optical and X-ray bands, $t_{\rm em,cj}$ are $\sim$ 3000 days and $\sim$ 5500 days post-merger, respectively.
From the left panel of Figure \ref{fig1}, it can be seen that
when the physical parameters are the same for the two branches,
the counter-jet component is generally very weak as compared with the emission from the near-jet.
At 3 GHz, the peak flux density of the counter-jet component is $8.5 \times 10^{-2} \, \mu \rm Jy$,
which is lower by $\sim$ 2 times than that of the near-jet at the same moment.
In X-ray band, the counter-jet plays an even minor role. The ratio of the peak flux of
the counter-jet to the near-jet flux at the same moment is about $1/4$.
However, since the counter-jet component generally has a slower declining rate after
the peak time, it tends to play a more and more significant role at late stages.
As a result, the afterglow light curves obviously become flatter after about $10^4$ days.
Thus although the counter-jet component is generally very weak, it still could be hopefully
detected through high accuracy observations of the late afterglow.
In this aspect, radio wavelength will be a more preferred window for the operation, as could be seen from our plot.
In Figure 1, we have compared our results with the 1$\sigma$ detection limits of several
large radio facilities, such as the Very Large Array \citep[VLA,][]{Perley2011},
the Square Kilometer Array \citep[SKA;][]{Dewdney2009},
and the Five-hundred-meter Aperture Spherical radio Telescope \citep[FAST;][]{Nan2011,Li2013}.
The 1$\sigma$ sensitivities are calculated at a representative band of 3 GHz
by using Equation (9) of \citet{Zhang2015}, assuming an integration time of one hour.
We see that the radio plateau can be hopefully revealed by FAST and SKA.

On the other hand, it is interesting to note that some physical parameters
of the twin jets may be different \citep{Huang2004,Jin2007,Racusin2008}.
In this study, we also explore how these possibilities could affect the role of the counter-jet.
First, let us consider the case that the ambient medium densities of the two branches are different.
As an example, under the assumption of $n_{\rm cj}$=100 $n_{\rm nj}$,
we have re-calculated the theoretical afterglow light curves. The results are
shown in the right panel of Figure \ref{fig1}.
Analytically, a larger number density of the ambient environment can lead to
a faster deceleration of the jet, therefore, an earlier peak time and a stronger emission.
As expected, at 3 GHz, the peak time of the counter-jet component is $ \sim 1500$ days,
and the corresponding peak flux is as large as a few $\mu$Jy, which is even higher
than the contribution from the near-jet at the same time,
and can be obtained successfully by FAST and SKA.
More encouragingly, after the peak time, the slope of the counter-jet component is much
flatter than that of the near-jet component, so that the emission from the counter-jet
completely dominate over that of the near-jet in radio and optical bands.
In fact, a clear plateau could be observed in the radio and optical light curves after 600 --- 1000 days.
At X-ray band, we see that the counter-jet component is still very weak.
It peaks at $\sim$ 2000 days, and could only marginally enhance the emission even at very late stages.

Figure \ref{fig2} illustrates the effect of the viewing angle $\theta_{\rm v}$ on the
afterglow light curve. The viewing angle is defined as the angle between our
line of sight and the near-jet axis.
It can be seen that when the viewing angle increases, the emission from the near-jet
is significantly reduced, especially at early stages. When $\theta_{\rm v}=90^{\rm o}$,
the contributions from the two branches are actually equal, assuming identical jets.
A larger viewing angle thus makes it easier to detect the counter-jet component.

In Figure \ref{fig3}, we show our results for the cases that various physical parameters
are different for the two branches of the double sided jet.
In each panel, we only change one parameter for the counter-jet, with all other parameters
unaltered with respect to those in Figure~(\ref{fig1}a).
As can be seen in Figure (\ref{fig3}a) and Figure (\ref{fig3}b), a larger $\epsilon_B$ or $\epsilon_e$
for the counter-jet can significantly enhance its emission.
Note that for the case of $\epsilon_{B, \rm cj}= 100\, \epsilon_{B, \rm nj}$,
the counter-jet component can even clearly show up as an obvious rebrightening  (up to a few $\mu \rm Jy$)
at about 5000 --- 6000 days in the afterglow light curve.
Similarly, when we set $\epsilon_{e, \rm cj}= 10\, \epsilon_{e, \rm nj}$,
a marked plateau of several $\mu \rm Jy$ emerges in the light curve.

In Figure (\ref{fig3}c) and (\ref{fig3}d), we explore the effect of the isotropic energy ($E_{c,\rm cj}$)
and the half-opening angle ($\theta_{c, \rm cj}$) of the counter-jet's inner core, respectively.
The increase of $E_{c, \rm cj}$ can significantly enhance the counter-jet emission.
When $E_{c, \rm cj}$ is taken as 10 $E_{c, \rm nj}$, the counter-jet emission
can manifest as a significant rebrightening that begins to show up after $\sim$ 3000 days.
In this case, the counter-jet emission will be easily detected.
As for the parameter of $\theta_{c, \rm cj}$, we see that its effect on the
emission is not significant.

According to \citet{Li2004}, the counter-jet emission peaks at the time when the ejecta becomes
non-relativistic, i.e., $t_{\rm NR,cj} \approx t_{\rm peak,cj} = (5/2)(1+z)(3E_{\rm cj}/4\pi m_p c^5 n_{\rm cj})^{1/3}$,
under the condition that the lateral velocity is zero.
For the cases of $n_{\rm cj} = n_{\rm nj}$ and $n_{\rm cj} = 100\, n_{\rm nj}$,
it can then be estimated that $t_{\rm NR,cj} \sim 5.4 \times 10^{3}$ days and $\sim 1.2 \times 10^{3}$ days, respectively.
Our numerical results are generally consistent with these analytical timescales. Of course, in our modeling,
the exact peak time is also slightly modified by the equal arrival time surface effect \citep{Wang2009,WangH2010,Geng2016}.
In our current semi-analytic calculations, similar to what has been done by \citet{Ghirlanda2019},  
we do not take into account the effect of lateral expansion, inclusion of which will need some elaborate considerations.
As a result, our modeling could be a conservative estimation.
The lateral expansion becomes significant when the jet has a Lorentz
factor $\sim$ 2---3 \citep{vanEerten2010,vanEerten2012a,vanEerten2012b,Granot2012}. 
Note that an approximation for lateral expansion is included for structured jets by \citet{Lamb2018}, following 
the method of \citet{Lamb2018a}. Their consideration is based on the fact that the lateral expansion predominantly 
affects the change in radius as the blast-wave expands, so that including this effect in calculating the equal arrival 
time surface is a good approximation for the sideways expansion of a structured jet. Additionally, the lateral 
spread due to sideways expansion is small when compared to the radial distance of the blast-wave.
Without lateral expansion, the deceleration of the jet is slower, so that the counter-jet component appears
slightly later and is also less significant than that in \citet{Wang2009}, \citet{vanEerten2011} and \citet{Granot2018}.
According to \citet{Zhang2009}, at the peak time of the counter-jet, the ratio of its flux over
the emission of the near-jet can be estimated
as $F_{\nu, {\rm cj}}/F_{\nu, {\rm nj}} \approx (1/3)^{(21-15p)/10} \approx 3.5 $ for GW170817/GRB 170817A.
Thus our semi-analytic calculations may underestimate the emission of the counter-jet component to some extent.

\section{Discussion and Conclusion}

GW170817/GRB 170817A, located at a luminosity distance of $D_{\rm L}= 40$ Mpc
\citep{Abbott2017b,Hjorth2017}, has been monitored at various bands for more than one year.
This provides us a good opportunity to investigate the afterglow of a double-sided jet
launched by the central engine.
In this study, we calculate the long time afterglow from the binary neutron star merger event numerically,
paying special attention to the counter-jet component.
It is found that its peak flux density is
$\sim 8.5 \times 10^{-2}\,\mu$Jy at 3 GHz, which is about 2 times lower than the corresponding emission from
the near-jet at the same time.
The existence of the counter-jet component makes the decay of the late time afterglow
much slower and makes the late time light curve significantly flatter, thus could potentially be observed by
large radio telescopes such as FAST and SKA.
At X-ray bands, the counter-jet component is even weaker and is essentially undetectable.

In our calculations, we have also considered the cases that the physical parameters of the two jet branches are different.
The difference could be resulted from a few factors.
For example, at a late stage ($>$ 100 --- 1000 days), the jet head would reach a radius of $\sim$ 1 pc
from the central engine, which is larger than the scale of normal Oort cloud.
The inhomogeneity of the environment medium may exist on this large scale.
Moreover, the double neutron star system will merger only after a long time of spiral-in ($10^6$ --- $10^7$ years).
During this process, the binary may drift in its host galaxy and may finally be far away from its birth place.
It can also lead to a different environment for the two jet branches.
Assuming that the ambient medium density of the counter-jet is 100 times higher
than that of the near-jet, the contribution from the
counter-jet is greatly enhanced. It can show up as a clear plateau about 600 days after the burst.
Similarly, if a larger value is assigned to the micro-physical parameters such as
$\epsilon_{B, \rm cj}$, $\epsilon_{e, \rm cj}$, and $E_{c, \rm cj}$ of the counter-jet,
then the counter-jet component will also be significantly enhanced.
It may manifest as an obvious plateau or even a rebrightening.
We thus argued that the observation of the very late afterglow of GW170817/GRB 170817A
can help to constrain the micro-physical parameters of the event.

In our study, for simplicity, we have investigated the cases that only one single parameter
is different for the near-jet and the counter-jet (see our Figures \ref{fig1} and \ref{fig3}).
However, in reality, it is also possible that these effects may be superposed.
For example, while the density of the circum-burst medium is high for the counter-jet,
the energy ratio of magnetic field (i.e. $\epsilon_{B}$) may also be high at the same time.
In this case, the emission from the counter-jet will be even stronger and may be easily detected.
Other parameters such as $\epsilon_{e, \rm cj}$ and $E_{c, \rm cj}$ are similar and we should
keep an open mind on these possibilities.

The circum-burst medium density of short GRBs
is generally as low as $\sim 10^{-3}$ --- $10^{-2}\rm\, cm^{-3}$
\citep{Fong2015,Hallinan2017}. For GRB 170817A, the number density of its surrounding medium has
also been determined as being in the range of $3\times 10^{-4}$ --- $2.4 \times 10^{-2}\rm\, cm^{-3}$.
However, strictly speaking, this density can only be regarded as the density of the medium
surrounding the near-jet. As for the density of the medium around the counter-jet,
we must resort to the very late afterglow, i.e. the counter-jet component.
As pointed out by \citet{Zhang2009},
$t_{\rm em,cj}$ is relevant to $E_{\rm cj}$ and $n_{\rm cj}$.
So, if the plateau or even a rebrightening is detected successfully,
we can constrain the circum-burst density of the counter-jet as
$n_{\rm cj} = 1.0 \times 10^{-2} (1+z)^3 E_{\rm cj,51} t_{\rm em,cj,3}^{-3} \rm\, cm^{-3}$.
Note that if the effect of lateral expansion is considered, the peak time of the counter-jet component is also relevant to
$\theta_{\rm cj}$. According to \citet{Li2004}, the circum-burst density can be constrained as
$n_{\rm cj} = 1.4 \times 10^{-3} (1+z)^3 E_{\rm cj,51} \theta_{\rm cj,-1}^{2} t_{\rm NR,cj,3}^{-3} \rm \, cm^{-3}$.
It will be helpful for us to acquire a thorough knowledge of the overall environment of the binary neutron star merger event.

It is interesting to note that late-time radio emission can also arise from
the dynamical ejecta launched during the merging process \citep[e.g.,][]{Hoto2016,Hoto2018}, but it
could be discriminated from the counter-jet emission through long-time followup observations.
The fast tail of the dynamical ejecta, with a smaller mass and a mildly-relativistic velocity,
is expected to emit isotropically through synchrotron mechanism and dominate the early light curve.
It can be examined by very long baseline interferometry observations \citep{Mooley2018x}.
The slow dynamic ejecta, with a larger mass and a sub-relativistic velocity,
will generate a long-lasting component which peaks at around $10^4$---$10^5$ days in the radio light curve.
In our model, the peak time of the counter-jet emission is earlier than that of the slow dynamical ejecta.

In short, the observation of the counter-jet can provide a lot of information on GRB outflows,
and can help to reveal the micro-physics of external shocks and the circum-burst environment.
It is thus necessary to continuously monitor the afterglow of GRB 170817A to a very late stage.
Emission from the counter-jet may hopefully be detected 600 --- 1000 days after the merger,
which is consistent with the statements of \citet{Gill2018} and \citet{Lamb2018}.
We expect that a positive detection of the afterglow at these late stages can be
obtained by FAST and SKA.

\acknowledgments
We thank the anonymous referee for constructive suggestions
that help to improve this study significantly.
This work is partially supported by the National Natural Science Foundation of China
(Grants No. 11873030 and 11833003), and by the Strategic Priority Research
Program of the Chinese Academy of Sciences ``Multi-waveband Gravitational Wave Universe''
(Grant No. XDB23040000).
BL acknowledges support from the National Program on Key Research and Development Project (Grant No. 2016YFA0400801)
and the Joint Funds of National Natural Science Foundation (Grant No. U1838113).



\clearpage

\end{document}